\documentstyle[aps,epsfig,preprint,floats]{revtex}
\begin{document}
\tighten

\def\lsi{\raise0.3ex\hbox{$<$\kern-0.75em\raise-1.1ex\hbox{$\sim$}}}
\def\gsi{\raise0.3ex\hbox{$>$\kern-0.75em\raise-1.1ex\hbox{$\sim$}}}

\renewcommand\({\left(}
\renewcommand\){\right)}
\renewcommand\[{\left[}
\renewcommand\]{\right]}

\newcommand\del{{\mbox {\boldmath $\nabla$}}}
\def\dbibitem#1{\bibitem{#1}\hspace{1cm}#1\hspace{1cm}}
\newcommand{\dlabel}[1]{\label{#1} \ \ \ \ \ \ \ \ #1\ \ \ \ \ \ \ \ }
\def\dcite#1{[#1]}

\newcommand\eq[1]{Eq.~(\ref{#1})}
\newcommand\eqs[2]{Eqs.~(\ref{#1}) and (\ref{#2})}
\newcommand\eqss[3]{Eqs.~(\ref{#1}), (\ref{#2}) and (\ref{#3})}
\newcommand\eqsss[4]{Eqs.~(\ref{#1}), (\ref{#2}), (\ref{#3})
and (\ref{#4})}
\newcommand\eqssss[5]{Eqs.~(\ref{#1}), (\ref{#2}), (\ref{#3}),
(\ref{#4}) and (\ref{#5})}
\newcommand\eqst[2]{Eqs.~(\ref{#1})--(\ref{#2})}

\newcommand\pa{\partial}
\newcommand\pdif[2]{\frac{\pa #1}{\pa #2}}

\def\be{\begin{equation}}
\def\ee{\end{equation}}
\def\ba{\begin{eqnarray}}
\def\ea{\end{eqnarray}}

\newcommand{\labell}[1]{\label{#1}\qquad_{#1}} 
\newcommand{\labels}[1]{\vskip-2ex$_{#1}$\label{#1}} 
\newcommand\gapp{\mathrel{\raise.3ex\hbox{$>$}\mkern-14mu
              \lower0.6ex\hbox{$\sim$}}}
\newcommand\gsim{\gapp}
\newcommand\gtsim{\gapp}
\newcommand\lapp{\mathrel{\raise.3ex\hbox{$<$}\mkern-14mu
              \lower0.6ex\hbox{$\sim$}}}
\newcommand\lsim{\lapp}
\newcommand\ltsim{\lapp}
\newcommand\M{{\cal M}}
\newcommand\order{{\cal O}}

\newcommand\extra{{\rm {extra}}}
\newcommand\FRW{{\rm {FRW}}}
\newcommand\brm{{\rm {b}}}
\newcommand\ord{{\rm {ord}}}
\newcommand\Pl{{\rm {pl}}}
\newcommand\Mpl{M_{\rm {pl}}}
\newcommand\mgap{m_{\rm {gap}}}
\newcommand\gB{g^{\left(\rm \small B\right)}}

\def\smallfrac#1#2{\hbox{${\scriptstyle#1} \over {\scriptstyle#2}$}}
\def\fourth{{\scriptstyle{1 \over 4}}}
\def\half{{\scriptstyle{1\over 2}}}
\def\st{\scriptstyle}
\def\sst{\scriptscriptstyle}
\def\mco{\multicolumn}
\def\epp{\epsilon'}
\def\vep{\varepsilon}
\def\ra{\rightarrow}
\def\ppg{\pi^+\pi^-\gamma}
\def\vp{{\bf p}}
\def\ko{K^0}
\def\kb{\bar{K^0}}
\def\al{\alpha}
\def\ab{\bar{\alpha}}

\def\calm{{\cal M}}
\def\calp{{\cal P}}
\def\calr{{\cal R}}
\def\calpr{{\calp_\calr}}

\newcommand\bfa{{\bf a}}
\newcommand\bfb{{\bf b}}
\newcommand\bfc{{\bf c}}
\newcommand\bfd{{\bf d}}
\newcommand\bfe{{\bf e}}
\newcommand\bff{{\bf f}}
\newcommand\bfg{{\bf g}}
\newcommand\bfh{{\bf h}}
\newcommand\bfi{{\bf i}}
\newcommand\bfj{{\bf j}}
\newcommand\bfk{{\bf k}}
\newcommand\bfl{{\bf l}}
\newcommand\bfm{{\bf m}}
\newcommand\bfn{{\bf n}}
\newcommand\bfo{{\bf o}}
\newcommand\bfp{{\bf p}}
\newcommand\bfq{{\bf q}}
\newcommand\bfr{{\bf r}}
\newcommand\bfs{{\bf s}}
\newcommand\bft{{\bf t}}
\newcommand\bfu{{\bf u}}
\newcommand\bfv{{\bf v}}
\newcommand\bfw{{\bf w}}
\newcommand\bfx{{\bf x}}
\newcommand\bfy{{\bf y}}
\newcommand\bfz{{\bf z}}

\newcommand\sub[1]{_{\rm #1}}
\newcommand\su[1]{^{\rm #1}}

\newcommand\supk{^{(K) }}
\newcommand\supf{^{(f) }}
\newcommand\supw{^{(W) }}
\newcommand\Tr{{\rm Tr}\,}

\newcommand\ncobe{N\sub{COBE}}
\newcommand\vev[1]{\langle{#1}\rangle}
\newcommand\mpl{M_{\rm P}}
\newcommand\GeV{\,\mbox{GeV}}
\newcommand\TeV{\,\mbox{TeV}}
\newcommand\MeV{\,\mbox{MeV}}
\newcommand\eV{\,\mbox{eV}}
\newcommand\mpsis{|m^2|}
\newcommand\etapsi{\eta_\phi}
\newcommand\sigmacob{\sigma\sub{COBE}}
\newcommand\eea{\end{eqnarray}}
\newcommand\bea{\begin{eqnarray}}

\title{ Hybrid Inflation and Baryogenesis at the TeV Scale}

\author{
Edmund J. Copeland$^1$, David Lyth$^2$, Arttu Rajantie$^3$ and
Mark Trodden$^4$}

\address{~\\
$^1$ Centre for Theoretical Physics,
University of Sussex,
Falmer, Brighton BN1 9QJ, UK \\
$^2$ Physics Department,
Lancaster University,
Lancaster, LA1 4YB, UK \\
$^3$ DAMTP, CMS, University of Cambridge,
Wilberforce Road,
Cambridge, CB3 0WA, UK \\
$^4$ Department of Physics,
Syracuse University,
Syracuse, NY, 13244-1130 USA}

\maketitle

\begin{abstract}
We consider the construction of inverted hybrid inflation models in which
the vacuum energy during inflation is at the TeV scale, and the
inflaton couples to the Higgs field.
Such models
are of interest
in the context of some recently proposed
models of electroweak baryogenesis.
We demonstrate how constraints on these models arise from quantum corrections,
and how self-consistent examples may be constructed, albeit
at the expense of fine-tuning. We discuss
two possible ways in which
the baryon asymmetry of the universe may be produced
in these models. One of them is based on preheating and a consequent
non-thermal electroweak symmetry restoration and
the other on
the formation of Higgs winding configurations by the Kibble mechanism
at the end of inflation.\\
\\
SUSX-TH/01-014, DAMTP-2001-18, SU-GP-01/2-3
\end{abstract}

\section{Introduction}
\label{introduction}
There is overwhelming observational evidence (for recent reviews see, e.g.,
\cite{Rubakov:1996vz,Trodden:1999ym,Riotto:1999yt})
that the universe has a
non-zero baryon density, quantified by
\begin{equation}
\frac{n_B}{s} \sim 10^{-10} \ ,
\end{equation}
where $n_B$ is the 
baryon number density in the universe, and $s$ is the entropy
density.
Because a period of cosmological inflation
\cite{Guth:1981zm,linde,a&s}
would have diluted away any baryons, this asymmetry must have been generated
afterwards.
One attractive possibility is
that baryogenesis proceeds through anomalous electroweak processes
occurring during the
departure from equilibrium provided by the electroweak phase
transition~\cite{Kuzmin:1985mm}.
However, the transition must be strongly first order
for electroweak sphaleron processes not to wash away the
baryon asymmetry immediately after it is generated,
and lattice Monte Carlo simulations have shown that this is not the
case in the minimal standard model~\cite{Kajantie:1996mn}.

As an alternative to the above standard picture
of electroweak baryogenesis, it has recently been
suggested~\cite{Krauss:1999ng,Garcia-Bellido:1999sv} that the washout
problem could be avoided if inflation
ends at the electroweak scale, with the inflaton strongly coupled to
the Higgs field. Preheating~\cite{KLS 94,STB 95},
i.e., parametric resonance between the Higgs field and the inflaton,
would lead to a non-equilibrium state in which the baryon asymmetry
could be generated, and a low enough reheat temperature would
guarantee that once the system has thermalized, sphaleron processes
are too rare to wash it out.
This idea has the attractive feature that it
may predict new
phenomena, testable in collider experiments, and thus provide a
hope of a direct
probe into the physics of the early universe.

Inflation model-building is however beset with difficulties, which become
more severe as the inflation scale is decreased \cite{Lyth:1999xn}.
One problem
is the difficulty in arranging the necessary initial conditions for
inflation to begin~\cite{Lin90,LinLinSec5,LinLinMez,V&T}. Another problem
is to understand the extreme
flatness of the potential, which requires that every tree-level and every
loop-correction  term in the potential be small unless there are cancellations
between them. Turning to the case at hand, the most obvious way of achieving
inflation with the inflaton strongly coupled to the Higgs field is to
invoke the usual hybrid inflation model, in which the inflaton is
rolling towards the origin. This paradigm,
adopted so far \cite{Krauss:1999ng,Garcia-Bellido:1999sv}, is unfortunately
spoiled by the loop correction~\cite{Lyth:1999ty}; its strong logarithmic
variation prevents inflation, and cannot be cancelled by a reasonable number
of tree-level terms. In this paper,  we consider
the alternative paradigm of inverted hybrid inflation, where the field is
rolling away from the origin. In contrast with the usual case, the loop
correction can now be cancelled accurately by a suitable choice of just
the renormalizable tree--level terms,
leading to a viable, albeit fine-tuned, model which can
give successful baryogenesis.

The outline of the paper is as follows. In Section \ref{baryogenesis},
we discuss two different
ways that baryon asymmetry can be generated.
One of these is the
scenario put forward in
Refs.~\cite{Krauss:1999ng,Garcia-Bellido:1999sv}, involving resonant
production of sphalerons and Higgs winding configurations
during preheating, and the other is based on the Kibble
mechanism~\cite{Kibble:1976sj} and
is a variant of the mechanism discussed in
Refs.~\cite{Turok:1990in,Turok:1991zg}.
In Section~\ref{problems}, we recall the loop correction
problem encountered with  ordinary
hybrid inflation, and in Section~\ref{model}, 
we see how it can be avoided by
inverted hybrid inflation if the mass and quartic
self-coupling are fine-tuned to cancel the loop correction.
 In Section~\ref{cosmological},
we determine the region of the parameter space in which
our model satisfies the constraints arising from inflation and baryogenesis.
Finally, we summarize the essential
results of this paper and offer some comments on the role of
low-scale inflation models.

\section{Baryogenesis after TeV-Scale Inflation}
\label{baryogenesis}
In the electroweak theory, baryon number is linked to the Chern-Simons
number of the SU(2) gauge field
\begin{equation}
N_{\rm CS}= {g^2 \over 32 \pi^2} \int_0^t dt\int d^3 x
\epsilon^{\mu\nu\rho\sigma} {\rm Tr} F_{\mu\nu} F_{\rho\sigma}
\label{cs_number}
\end{equation}
by a quantum anomaly~\cite{'tHooft:1976up},
\begin{equation}
\Delta B=3\Delta N_{\rm CS}.
\end{equation}
In order to generate the baryon asymmetry, it is therefore necessary
that the dynamics of the system leads to a non-zero Chern-Simons
number.

The dynamics of the Chern-Simons number is linked to the dynamics of
the Higgs field via the Higgs winding number
\begin{equation}
N_H = \frac{1}{24\pi^2} \int d^3x\, \epsilon^{ijk} \hbox{Tr}
[U^{\dagger}\partial_iUU^{\dagger}\partial_jUU^{\dagger}\partial_kU].
\label{higgswinding}
\end{equation}
In this parameterization, the $SU(2)$ Higgs field $\Phi$ has been expressed as
$\Phi = ({\rho}/{\sqrt{2}}) U  $, where $\rho^2 =
2\left(\varphi_1^*\varphi_1 + \varphi_2^*\varphi_2 \right) = {\rm Tr} \Phi^\dagger
\Phi$,  and $U$ is an $SU(2)$-valued matrix that is uniquely defined
anywhere $\rho$ is nonzero.
In the broken phase
we have in practice
$N_{\rm CS}=N_H$, and thus any difference in the two numbers must
disappear when the system thermalizes.
This may either happen by changing the Chern-Simons number,
which would lead to baryon
production~\cite{Turok:1990in,Turok:1991zg}, or by changing the Higgs
winding, in which case no baryons are produced.
CP-violation
affects the balance between these two ways, and leads to an asymmetry
in the final baryon number (see \cite{Lue:1997pr} for a detailed discussion of
the dynamics of winding configurations).

\subsection{Baryogenesis from Preheating}
\label{ssect:barypre}
One possibility for baryon production arises from
{\it reheating} after cosmological inflation.
Careful, non-perturbative studies \cite{Khlebnikov:1996mc,Prokopec:1997rr}
of the inflaton dynamics have demonstrated that
there may be a period of parametric resonance, prior to the usual scenario of energy
transfer from the inflaton to other fields. This phenomenon is
characterized by large amplitude, non-thermal excitations in both the
inflaton and coupled fields, and has become known as {\it preheating}
\cite{KLS 94,STB 95}.

In the models we consider in this paper, the inflaton is directly
coupled to the standard
model Higgs field. Therefore, if preheating occurs, we expect long-wavelength
excitations of standard model fields to be resonantly produced.
This may lead to a non-thermal restoration of the SU(2)
symmetry~\cite{kofm,Tkachev:1996md,Rajantie:2000fd},
and to the production of a population of winding
configurations \cite{Krauss:1999ng} and a non-zero Chern-Simons number
\cite{Garcia-Bellido:1999sv} even though the
reheat temperature after inflation is never above the electroweak
scale,
and even
though the electroweak symmetry remains broken after the end of
inflation.
Because of the non-perturbative nature of this mechanism, it is very
difficult to derive reliable analytical estimates for the baryon
asymmetry, and one has to resort to numerical
simulations~\cite{Rajantie:2000nj}.

\subsection{Baryogenesis from the Kibble Mechanism}
\label{ssect:barykibble}
The mechanism of the previous section; preheating and a subsequent
non-thermal symmetry
restoration, needs an initial energy density that is much higher
than the Higgs potential
barrier. However, if the energy density is lower, there is another way the
baryon asymmetry can be generated. This is based on the instability of
the Higgs field (i.e.,
tachyonic preheating \cite{Garcia-Bellido:1998wm,Felder:2000hj}), 
which leads to formation of topological 
``defects'' by
the Kibble
mechanism~\cite{Kibble:1976sj} when the SU(2)
symmetry breaks.

In hybrid inflationary models,
the Higgs field vanishes during inflation, and the
electroweak symmetry is therefore unbroken. At the end of inflation,
the Higgs field rolls to its true vacuum
\cite{Garcia-Bellido:1998wm,Felder:2000hj}.
The Kibble mechanism ensures that Higgs winding configurations
will form, and CP violation inherent in many extensions of the
standard model ensures that net baryon production results.
Because the universe is cold after the inflation, the phase
transition takes place at zero temperature. This means that there are
no sphaleron processes that would wash out the baryon asymmetry, and
consequently no need for a first-order phase transition. Of course,
the universe eventually reheats to the temperature $T_{\rm rh}$,
but if this is low enough so that the electroweak symmetry is strongly
broken, the baryon asymmetry
is safe.

The baryon number generated in this scenario can be estimated by
finding the shortest wavelength that falls out of
equilibrium~\cite{Zurek:1993ek}. When
the inflaton field $\sigma$ rolls toward its minimum,
the effective mass squared of the Higgs field
$m^2(\sigma)$ changes, and at $\sigma=\sigma_c$ it changes
sign.
Assuming that a Fourier mode with momentum $k$ is in equilibrium, its
frequency is given by
\begin{equation}
\omega(k)^2 = k^2+m^2(\sigma).
\label{momentum}
\end{equation}
If the equation
\begin{equation}
\left|\frac{d\omega(k)}{dt}\right| \gsim \omega(k)^2
\label{adiabaticity}
\end{equation}
is satisfied, the mode behaves non-adiabatically, i.e., it evolves so
slowly that it does not have time to adjust to the change of its
effective mass. Consequently, the mode falls out of equilibrium. In
general, there is a critical momentum $k_{\rm max}$ such that this
happens for $k\le k_{\rm max}$.
The maximum correlation length
reached during the transition is given
by $\xi_{\rm max}=k_{\rm max}^{-1}$.

When $m^2(\sigma)$ becomes negative at $\sigma_c$, the Higgs field
acquires a non-zero expectation value, but its direction can only be
correlated at distances less than $\xi_{\rm max}$. This leads to
domains of radius $\xi_{\rm max}$, inside each of which the Higgs
field is roughly constant,
and when
the field is interpolated between these domains, it typically acquires
a non-zero winding number. The resulting number density of these winding
configurations is determined by the domain size,
\begin{equation}
n_{\rm configs}
\sim \xi_{\rm max}^{-3} \sim k_{\rm max}^3,
\end{equation}
and a rough estimate indicates that this density must be higher than
$0.001m_H^3$ for sufficient baryogenesis~\cite{Krauss:1999ng}.

The above analysis ignores the gauge field, which plays an important
role in defect formation in the high-temperature 
case~\cite{Hindmarsh:2000kd}. The reason why this is justified here is
that
the effect of the gauge field is proportional to the temperature, and 
in our case, the symmetry breaking takes place at zero temperature.

\section{Constraints on Hybrid Inflation Models}
\label{problems}
\subsection{Basics}
In order to analyse the viability of different models of TeV-scale inflation,
let us first
summarize the basics of inflation model building, as given in say
\cite{Lyth:1999xn}.
We write the reduced Planck mass as $\mpl\equiv (8\pi G)^{-1/2}
=2.4\times 10^{18}\GeV$, and the Hubble parameter as $H$.
Further, an overdot denotes
differentiation with respect to time and the prime differentiation with
respect to the inflaton field $\sigma$.

We will be interested in the slow-roll paradigm of inflation
in which the field equation for the inflaton $\sigma$
\begin{equation}
\ddot\sigma+3H\dot\sigma+V'=0 \ ,
\end{equation}
is replaced by the slow-roll condition
\begin{equation}
\label{equ:slowroll}
3H\dot\sigma\simeq-V' \ .
\end{equation}
We also require
\begin{equation}
V \simeq \rho \simeq 3\mpl^2H^2 \ ,
\end{equation}
with $\rho$ the energy density, be almost constant on the Hubble timescale.
The latter condition requires one flatness condition
\be
\epsilon \ll 1 \label{flat1} \,,
\ee
and differentiating the slow-roll condition requires
a second flatness condition
\be
|\eta| \ll 1 \label{flat2}
\,,
\ee
where
\bea
\epsilon&\equiv &\frac12\mpl^2 \left(\frac{V'}{V}\right)^2 \,,\\
\eta&\equiv &\mpl^2 \frac{V''}{V} \,.
\eea
$N$ $e$-folds before the end of slow-roll inflation, the field value
$\sigma_N$ is given by
\be
N = \frac{1}{\mpl^{2}}\int^{\sigma_N}_{\sigma_{\rm end}} \frac V{V'} d\sigma \,,
\label{nint}
\ee
where $\sigma\sub{end}$ marks the end of slow-roll inflation.

If the inflaton field fluctuation is responsible for structure in the
Universe, the COBE measurement of the cosmic microwave background anisotropy
requires
\be
\frac{V(\sigma_{\rm COBE})^{3/2}}
{\mpl^{3}V'(\sigma_{\rm COBE})} = 5.3\times 10^{-4} \,.
\label{cobenorm}
\ee
This equation applies at the epoch when the distance scale
explored by COBE (say $H_0^{-1}/10$) leaves the horizon,
some number $N\sub{COBE}<60$ $e$-folds before the end
of slow-roll inflation.

\subsection{Ordinary hybrid inflation}
In Refs.~\cite{Krauss:1999ng,Garcia-Bellido:1999sv}, baryogenesis was
discussed in the context of ordinary hybrid inflation.
In this model, the tree-level potential is \cite{andhyb}
\be
V(\sigma,\phi)= V_0 + \Delta V(\sigma)
-\frac12\mpsis\phi^2 +\frac12{g^2}\phi^2\sigma^2
+ \frac14\lambda\phi^4 \,,
\label{fullpot}
\ee
where, in the simplest model,
\be
\Delta V(\sigma) = \frac12m_\sigma^2\sigma^2 \,.
\label{mterm}
\ee
Inflation takes place in the regime $\sigma^2>\sigma\sub c^2$, where
\be
\sigma\sub c \equiv \frac{|m|}{g} \,.
\label{phic}
\ee
In this regime, $\phi$ vanishes
and the inflaton potential is
\be
V=V_0+ \Delta V(\sigma) \ , \label{vofphi2}
\ee
where the constant term $V_0$ is assumed to dominate during inflation \cite{andhyb,copelandhyb}.

The specific implementation proposed in~\cite{Krauss:1999ng,Garcia-Bellido:1999sv}
involved the tree-level hybrid inflation
model using the Higgs as the non-inflaton field.
This same paradigm was suggested later
\cite{kl} (without specifically invoking the Higgs field)
as a model of inflation with quantum gravity at the TeV scale.
Unfortunately both of these proposals are spoiled by the loop correction
 \cite{Lyth:1999ty}.

To see how this comes about, let us first pursue the model ignoring the loop correction.
The  last term of
\eq{fullpot} serves only to determine the vacuum
expectation value (VEV) of $\phi$, achieved when $\sigma$ falls below
$\sigma\sub c$. Demanding that $V$ vanish in the vacuum, so that the cosmological
constant is zero after inflation, implies that
the VEV is
\be
\langle\phi\rangle\equiv
v = 2 \frac{V_0^{1/2}}{|m|} \,,
\label{mofv} \ee
and that
\be
\lambda=\frac{4V_0}{v^4} = \frac{m^4}{4V_0} \,.
\label{lamofv}
\ee
{}From \eq{mterm},
\be
\eta = \frac{m_\sigma^2 \mpl^2}{V_0} \,.
\ee
It will be useful also to define
\be
\etapsi \equiv \frac{\mpsis\mpl^2}{V_0} =\frac{4\mpl^2}{v^2}
\label{etapsiapp}
\,.
\ee
A prompt end to inflation at
$\sigma\sub c$ requires
\be
\etapsi \gsim  1 \,,
\ee
which is well satisfied.
Now, the value of the field $\sigma$ when COBE scales leave the horizon is
\be
\label{sigmacobe}
\sigmacob^2 =e^{2\eta N\sub{COBE}} \sigma\sub c^2 \simeq \sigma\sub c^2
\,,
\label{phi}
\ee
where the final equality is good for an order of magnitude estimate,
but needs to be checked for consistency in any particular example.
Using~(\ref{sigmacobe}), the
COBE normalization is
\bea
{g^2} &=& 2.8\times 10^{-7} e^{2\eta N\sub{COBE}} \eta^2 \eta_\phi \\
&\simeq& 3 \times 10^{-7} \eta^2 \eta_\phi \,.
\label{cobenorm2}
\eea
To identify $\phi$ with the Higgs field we need roughly  $m \sim
$\,TeV
and $\lambda\sim 1$ (corresponding to $V_0^{1/4}\sim v\approx 250 $\,GeV)
 and for baryogenesis to work we need a strong coupling to the Higgs
 field,
$g\sim 1$.
The COBE normalization then requires $m_\sigma\sim 10^{-10}\eV$,
which corresponds to
$\eta \simeq 10^{-15}$, justifying the final equality in \eq{phi}.
Because $\eta$ is so small, $\sigma$ is practically equal to the
critical value
$\sigma\sub c$ while scales of interest are leaving the horizon.
On the other hand,
it seems reasonable to require that
inflation occurs  over some reasonable range of $\sigma$,
since otherwise one would have to explain how $\sigma$ arrived at
precisely the value $\sigma\sub c$.

The  small value of $m_\sigma$ means that
even if the loop correction
did not spoil the model,  the energy density
at the end of inflation would be  too
low for the scenarios
discussed in Section~\ref{baryogenesis}.
Because of the slow-roll condition (\ref{equ:slowroll}),
the kinetic energy is negligibly small,
\begin{equation}
\frac{1}{2}\dot{\sigma}^2
=\frac{M_p^2V'^2}{6V}\sim\frac{M_p^2m_\sigma^4}{m_H^2}
\lsim (0.01~{\rm eV})^4.
\end{equation}
The energy density consists therefore only of the potential energy,
which
at $\sigma=\sigma_c$ is
$V(\sigma_c)=V_0+\frac{1}{2}m_\sigma^2\sigma_c^2\approx V_0$.
Because $V_0$ is the height of the Higgs barrier,
the energy density must be much higher than this for symmetry restoration
as discussed in Section~\ref{ssect:barypre} to be possible.
The low kinetic energy also means that the Kibble mechanism
discussed in Section~\ref{ssect:barykibble} cannot lead to
significant baryogenesis either.

Now consider the loop correction, coming from the Higgs field.
Part of this correction just renormalizes the mass and quartic self-coupling
of the inflaton field. If there is no supersymmetry (SUSY), one  has to
take the view that the renormalized couplings are set to desired values,
making this part of the loop correction insignificant.
If, on the other hand, there is
supersymmetry, this part of the
loop correction vanishes (in the global supersymmetric limit, with
soft or
spontaneous SUSY breaking, and after the Higgsino and the
other Higgs field have been
included).
The other, logarithmic, part of the loop correction is
more problematic. This contribution is
\be
\label{equ:loopcorr}
\Delta V\sub{loop}(\sigma)=
\frac{1}{64\pi^2}\(m^4(\sigma)
{\rm ln}\frac{m^2(\sigma)}{Q^2} \)
\,,
\ee
where
\be
m^2(\sigma) \equiv \({g^2}\sigma^2-\mpsis \)
={g^2}(\sigma^2-\sigma\sub c^2)
 \,.
\ee
The quantity $Q$ is the renormalization scale at which the parameters of the
tree-level potential should be evaluated. Its choice is arbitrary,
and if all loop corrections were included,
the total potential would be independent
of $Q$ by virtue of the Renormalization Group Equations (RGEs).
In any application of quantum field theory, one
should choose $Q$ so that the total 1-loop correction is
small, hopefully justifying the neglect of the multi-loop correction.

Unless $\sigma$ is extremely close to $\sigma\sub c$,
the loop correction and its derivatives are
roughly estimated by setting $\sigma \sub c=0$, giving
\bea
\Delta V\sub{loop} &=& \frac {g^4}{64\pi^2} \sigma^4 \ln 
\frac{g^2 \sigma^2}{Q^2}, \\
\Delta V\sub{loop}' &=& \frac {g^4}{32\pi^2} 
\sigma^3 \(2\ln \frac{g^2 \sigma^2}{Q^2}
+1\),\\
\Delta V\sub{loop}'' &=& \frac {g^4}{32\pi^2}\sigma^2 
\(6 \ln \frac{g^2 \sigma^2}{Q^2}
+ 7\).
\eea
Because of the logarithmic variation, the loop correction cannot be cancelled to high accuracy
over a reasonable range of $\sigma$, by a tree-level contribution
 containing a reasonable number
of terms. The flatness conditions $\epsilon\ll 1$ and $\eta\ll 1$ therefore both require
\be
g\ll \frac{v}{\mpl} \,.
\label{flatmarch}
\ee
(One of these can be avoided by a choice of $Q$, but not both.)
This constraint precludes a VEV $v$ at the TeV scale, except for an unfeasibly small value
of $g$. It
 holds independently of the form of the tree-level contribution. Taking the
simplest form \eq{mterm},  requiring  that the COBE normalization is not upset by the loop
correction gives an additional constraint \cite{Lyth:1999ty}
\be
v^4 \sigma\sub{COBE}
\gsim
 \( 10^{9}\GeV \)^5\( \frac{V_0^{1/4}}{1\MeV} \)^2 \,.
\label{fourth}
\ee
This
precludes, by many orders of magnitude, hybrid inflation with
$\sigma\sub{COBE}$, $v$ and $V_0^{1/4}$ at the TeV scale.

As we noted earlier, with TeV--scale inflation, 
length scales corresponding
to our observable Universe,
actually leave the horizon when $\sigma$ is very
close to $\sigma\sub c$. 
In principle, one could therefore avoid the above constraints by
starting the inflation very close to $\sigma_c$.
In that regime the loop correction is very suppressed, and so is
its first derivative making
$\epsilon$ very small. Although the second derivative is not suppressed in
general, 
it can
be suppressed at any single point
by a suitable choice of the renormalization scale $Q$. 
However, the third derivative $\Delta V_{\rm loop}'''$ will not be
suppressed at the same point, and near $\sigma_c$ it is roughly
\begin{equation}
\Delta V_{\rm loop}'''\sim \frac{g^4\sigma_c^2}{\sigma-\sigma_c}.
\end{equation}
Therefore, even if we tune $Q$ in such a way that the the slow-roll condition
(\ref{flat2}) is satisfied at $\sigma$, the length of the 
range inside which it remains
satisfied is of order
\begin{equation}
\Delta\sigma\sim\frac{V_0(\sigma-\sigma_c)}{g^2|m^2|M_p^2}
\sim \frac{v^2}{M_p^2}(\sigma-\sigma_c)\approx
10^{-32}(\sigma-\sigma_c).
\end{equation}
This is far too short to produce any inflation at all in practice,
because Eq.~(\ref{nint}) implies that $N\approx
\Delta\sigma/\eta\sigma
\sim 10^{-17}(\sigma-\sigma_c)/\sigma$.
We conclude that the hybrid inflation paradigm is unviable
at the TeV scale, and turn now
to an alternative.

\section{Inverted hybrid inflation}
\label{model} Given the concerns discussed in the previous
section, we shall turn from ordinary hybrid inflation, in which
the inflaton rolls  towards the origin, and consider instead the
case of inverted hybrid inflation \cite{burt,ourinvert,steve}
where it rolls away from the origin, and inflation takes place at
small $\sigma$. This is achieved by giving $\Delta V$ a negative
slope, making the coupling between $\phi$ and $\sigma$ negative,
and making $m^2$ positive.

The effective potential we shall consider is
\begin{equation}
V(\phi,\sigma)=V_0 -{1\over p}\lambda_p\sigma^p+{1\over q}\kappa_q\sigma^{q}
-{1\over 2}g^2\sigma^2\phi^2+{1\over 2}m^2\phi^2+{1\over
4}\lambda\phi^4 \ ,
\label{potential}
\end{equation}
where all the parameters in the potential are positive semi-definite.
To have  a viable model we shall need to consider non-renormalizable
terms $q>p>4$, corresponding to dimensionful parameters $\lambda_p$ and
$\kappa_q$.

The tree-level mass term $m^2$ of the Higgs field is positive, and therefore
the electroweak symmetry is restored during inflation, when $\sigma$
is small. When $\sigma$ reaches the critical value $\sigma_c=m/g$, the
Higgs field becomes unstable and the symmetry gets broken. If we assume that
the time scale of the dynamics of the Higgs field is much faster than
that of the inflaton, we can calculate  that at any value of $\sigma$,
the Higgs field has the value
\begin{equation}
\phi^2=\phi_{\rm min}^2={g^2\sigma^2-m^2\over\lambda} \ .
\end{equation}
Thus the effective potential for the inflaton $\sigma$ is
\begin{equation}
\label{equ:effpot}
V_{\rm eff}(\sigma)=V_0-{m^4\over 4\lambda}+{1\over 2}\frac{g^2 m^2}{\lambda}\sigma^2
-\frac{g^4}{4\lambda}\sigma^4 -{1\over p}\lambda_p\sigma^p+{1\over
q}\kappa_q\sigma^{q} \ .
\end{equation}

The loop correction is given by the same expression
(\ref{equ:loopcorr}) as before.
At $\sigma\gg\sigma_c$, it behaves as 
$\Delta V_{\rm loop}(\sigma)\sim\sigma^4\ln\sigma$, which is of lower
order than the tree-level terms $\sigma^p$ and $\sigma^q$, and 
therefore we can ignore the loop correction when we study the dynamics
after inflation. During inflation, $\sigma<\sigma_c$, so that,
\be
m^2(\sigma)=
m^2 - g^2\sigma^2
=g^2(\sigma\sub c^2-\sigma^2 ).
\ee
An appropriate choice in this regime is $Q=m$, and with that choice
the loop correction
is a power series, equivalent to a tree-level contribution.
At sufficiently small $\sigma$, the series
 converges rapidly and we need keep only the renormalizable terms
(quadratic and quartic). In the language of effective field theory,
we have obtained an effective
 field theory valid at $\sigma\ll m$  by integrating out
the physics on the scale $m$. This is analogous to the  way
in which one might obtain the Standard Model by integrating out
GUT physics on the scale $M\sub{GUT}$.

In order to satisfy the slow-roll and COBE constraints, the potential
must be very flat at small $\sigma$, where inflation takes place.
As a result, at least the quadratic and quartic terms must be
cancelled, by the tree-level terms of the original potential.
Demanding a perfect cancellation,
we arrive at
the ``renormalized'' loop correction
\be
\label{equ:renloopcorr}
\Delta V\sub{loop}(\sigma)=
\frac{1}{64\pi^2}\[m^4(\sigma)
{\rm ln}\frac{m^2(\sigma)}{m^2} +g^2m^2\sigma^2
-\frac{3}{2}g^4\sigma^4
\]
.
\ee

To achieve slow-roll inflation the inflaton mass-squared  has to be
much less than $V_0/\mpl^2\sim 10^{-30} m^2\sim (10^{-4}\eV)^2$, which means
that the quadratic counter-term has to cancel the loop correction to an
accuracy of $10^{-30}$. Also, the  dimensionless
coupling of the quartic term has to be $\lsim 10^{-15}$ to achieve the
COBE normalization,
and  the quartic counter-term has to cancel the loop correction to this
accuracy. These two extremely  fine-tuned cancellations are the price we have
to pay, for the not inconsiderable prize of
inflation at the TeV scale with
an unsuppressed coupling to the Higgs field leading to viable
baryogenesis. {}From the effective field theory viewpoint, we have
renormalized the parameters of the
effective  small-$\sigma$ potential, in an analogous way to that in which
the parameters of the Standard Model effective theory are renormalized
if it is obtained from a non-supersymmetric GUT theory.

The next term in the power series representing the loop correction is
\begin{equation}
\label{equ:smallsigmacorr}
\Delta
V\sub{loop}(\sigma)=-\frac{1}{192\pi^2}\frac{g^6\sigma^6}{m^2},
\end{equation}
and we shall
check that this does not spoil the COBE normalization.

Now, the effective potential (\ref{equ:effpot}) is minimized when
\begin{equation}
\label{equ:minicond}
\kappa_q \sigma_{\rm min}^{q-2} -\lambda_p \sigma_{\rm min}^{p-2}
-\frac{g^4}{\lambda}\sigma_{\rm min}^2 +\frac{g^2 m^2}{\lambda} =0 \ .
\end{equation}
Assuming that $g^2$ is small, the mass of the Higgs field in this
minimum, which corresponds to the physical vacuum, is
\begin{equation}
\label{equ:higgsmass}
m_H^2=2(g^2\sigma^2_{\rm min}-m^2) \ .
\end{equation}
Because $m_H^2$ is an observable quantity, and $\sigma_{\rm min}$ is not, it is
useful to invert this relation and write
\begin{equation}
\label{equ:sigmamin}
\sigma^2_{\rm min}=\frac{1}{g^2}\left(m^2+\frac{1}{2}m_H^2\right) \ .
\end{equation}
By requiring $V(\sigma_{\rm min})=0$, i.e. that the
cosmological constant be zero
in the physical vacuum, and using (\ref{equ:sigmamin}), we may
express $V_0$ in terms of the other parameters as
\begin{eqnarray}
\label{equ:inirho}
V_0 &=& - \frac{m^2}{\lambda}\left(\frac{1}{2}-\frac{1}{q}\right)
\left(m^2+\frac{1}{2}m_H^2\right) + \frac{1}{\lambda}\left(\frac{1}{4}-
\frac{1}{q}\right) \left(m^2+\frac{1}{2}m_H^2\right)^2 \nonumber \\
&& +\frac{\lambda_p}{g^p}\left(\frac{1}{p}-\frac{1}{q}\right)
\left(m^2+\frac{1}{2}m_H^2\right)^{p/2} +\frac{m^4}{4\lambda} \ .
\end{eqnarray}
For future convenience, let us introduce the dimensionless parameters
\begin{equation}
\label{equ:params}
\alpha\equiv\frac{2m^2}{m_H^2},\qquad\beta\equiv\frac{\lambda\lambda_p}
{m_H^4}\sigma_{\rm min}^p.
\end{equation}
In terms of these parameters, the initial energy density Eq.~(\ref{equ:inirho})
is
\begin{equation}
\label{equ:inirho2}
\rho_{\rm init}=V(\phi=0,\sigma=0)=V_0=
\frac{m_H^4}{4\lambda}\left[
\frac{1}{4}-\frac{\alpha+1}{q}+4\frac{q-p}{pq}
\beta\right].
\end{equation}

Using Eq.~(\ref{equ:minicond}),
we can also express $\kappa_q$ in terms of $m_H^2$ as
\begin{eqnarray}
\label{equ:kappa}
\kappa_q &=& g^q \left(m^2+\frac{1}{2}m_H^2\right)^{1-q/2}
\left[\frac{m_H^2}{2\lambda}+ \frac{\lambda_p}{g^p}
\left(m^2+\frac{1}{2}m_H^2\right)^{p/2-1}
\right] \nonumber\\
&=&
\frac{g^qm_H^{4-q}}{\lambda}\left[
\frac{\alpha+1}{4}
+\beta\right]\left(\frac{\alpha+1}{2}\right)^{-q/2}.
\end{eqnarray}

\section{Cosmological Constraints}
\label{cosmological}

Having defined the model,
we now
turn to the constraints imposed by the requirement of a
successful cosmological evolution. In particular, we demand the generation
of COBE scale fluctuations in the microwave background during inflation, followed by
the generation of the observed baryon asymmetry, both occurring around the electroweak
scale.

We can constrain the exponent $p$ by
writing the COBE normalization \cite{grav2}
in terms of the VEV of $\sigma$, estimated from the first two
terms of \eq{potential} as
\begin{equation}
\sigma_{\rm min}\sim \left(\frac{V_0 p}{\lambda_p}\right)^{1/p} \ .
\end{equation}
For $p=4$, 5 and 6 one finds
\bea
\frac{\sigma_{\rm min}}{100\GeV}
&=& 10^3 \( \frac{V_0^\frac14}{100\GeV} \) \ \ \
(p=4), \\
\frac{\sigma_{\rm min}}{100\GeV}
&=& 2.5 \( \frac{V_0^\frac14}{100\GeV}\)^\frac65 \ \ \
(p=5), \\
\frac{\sigma_{\rm min}}{100\GeV}
&=& 0.063 \( \frac{V_0^\frac14}{100\GeV} \)^\frac43
\ \ \
(p=6).
\eea
With  $p=5$,
the VEV and the height of the potential can both be at the electroweak
scale. This is what we need for baryogenesis, giving also the nice
feature that the dimensionful couplings are at the electroweak scale.
With $p=4$, the VEV is far above the electroweak scale, and the
dimensionless coupling is very small. With $p=6$,
the VEV  is appreciably  below the electroweak scale and the dimensionful
coupling somewhat above it. Only $p=5$ and (marginally) $p=6$ provide
a viable model of inflation with all relevant quantities around the
electroweak scale.

In order to see more precisely
how the COBE data constrains the parameters, we
use Eq.~(\ref{nint}) to calculate the value $\sigma_{\rm COBE}$ at
which the fluctuations were generated.
 It turns out that slow-roll inflation
ends at $\sigma\ll\sigma\sub c$, which means that the COBE constraint is
practically the same as in the corresponding `new'  inflation model
(the first three terms of \eq{potential}  with
$V$ vanishing at the minimum).
At small $\sigma$, we can
approximate $V(\sigma)\approx V_0$ and
$V'(\sigma)\approx -\lambda_p\sigma^{p-1}$ so that
Eq.~(\ref{nint}) becomes
\begin{equation}
N(\sigma_{\rm COBE})\approx \frac{1}{p-2}\frac{V_0}{\lambda_pM_p^2}
\sigma_{\rm COBE}^{2-p},
\end{equation}
and Eq.~(\ref{cobenorm}) becomes   \cite{Lyth:1999xn}
\begin{equation}
\frac{V_0^3}{\lambda_p^2M_p^6}\sigma_{\rm COBE}^{2-2p}=
\frac{V_0^3}{\lambda_p^2M_p^6}\left[(p-2)N\frac{\lambda_pM_p^2}{V_0}
\right]^{\frac{2p-2}{p-2}}
\approx 2.7\times 10^{-7} \ .
\end{equation}
These yield
\begin{equation}
\label{equ:lambda}
\lambda_p\approx \left(5.2\times 10^{-4}\right)^{p-2}
\left[(p-2)N\right]^{1-p}\left(\frac{M_p}{\sqrt{V_0}}\right)^{p-4}
\end{equation}
and
\begin{equation}
\sigma_{\rm COBE}\approx 1.9\times 10^3 (p-2)N\frac{\sqrt{V_0}}{M_p}
\sim 10~{\rm eV},
\label{sigmacobebound}
\end{equation}
which means that unless $m^2$ is extremely small, the fluctuations
were generated well before $\sigma_c$. This is crucial, because it
means that we need not worry about
the ``renormalized'' loop correction
Eq.~(\ref{equ:renloopcorr}). {}From this expression,  the contribution to $V'$ is
\begin{equation}
\Delta V'_{\rm loop}(\sigma_{\rm COBE})\approx-\frac{g^6\sigma_{\rm
COBE}^5}{32\pi^2m^2}.
\end{equation}
Therefore loop corrections do not change the COBE fluctuations if
\begin{equation}
\label{equ:loopconstraint}
\delta_{\rm loop}\equiv
\frac{g^6}{32\pi^2\lambda_p}\frac{\sigma_{\rm COBE}^{6-p}}{m^2}\ll 1.
\end{equation}
We shall check that this holds for the parameter values that we propose.
(The model may work even if this inequality is not satisfied, but then
the analysis becomes more complicated, because we cannot neglect the
loop correction.) Finally, the flatness condition $\eta\ll 1$ is
\be
10^9 g^3 \frac{m}{\mpl} \ll 1 \,,
\ee
which is easily satisfied by all the parameter values we will consider.

Since we are mainly interested in low-scale inflation models as a
way of avoiding the problems of ordinary electroweak baryogenesis,
the reheat temperature $T_{\rm rh}$ to which the universe
eventually equilibrates must be lower than the electroweak
critical temperature. Otherwise, the electroweak phase transition
would take place just as in the standard Big Bang scenario. In
fact, $T_{\rm rh}$ must be even lower, because if it is too close
to the critical temperature, sphaleron processes would be so
frequent that they would wash out the baryon asymmetry generated
during the non-equilibrium stage. This can only be avoided if the
Higgs field has a large enough expectation value at the reheat
temperature, $\phi(T_{\rm rh})\gsim T_{\rm rh}$
\cite{Shaposhnikov:1986jp,Shaposhnikov:1987tw}. This translates
into
\begin{equation}
\label{equ:Trhbound}
T_{\rm rh}\lsim 150~{\rm GeV}.
\end{equation}
The corresponding constraint for the energy density is
\begin{equation}
\rho(T_{\rm rh})\approx \frac{\pi^2}{30}g_*T_{\rm rh}^4\lsim
10^{10}~{\rm GeV^4}\approx (320~{\rm GeV})^4.
\end{equation}
Because energy is conserved, the initial energy density $\rho_{\rm
init}=V_0$ in Eq.~(\ref{equ:inirho2}) must satisfy the same bound.
This may be written in terms of $\alpha$ and $\beta$ as
\begin{equation}
\frac{1}{4}-\frac{\alpha+1}{q}+4\frac{q-p}{pq}
\beta
\lsim 20.
\label{conserve}
\end{equation}

Another condition the potential must satisfy is that $\sigma_{\rm
min}$ actually is a minimum rather than a maximum, since
Eq.~(\ref{equ:minicond}) only guarantees that it is an extremum.
This means that we must require $V''_{\rm eff}(\sigma_{\rm
min})>0$, and since
\begin{equation}
\label{equ:Vdoubleprime}
V''(\sigma_{\rm min})=\frac{g^2m_H^2}{2\lambda}\left[
(2-q)\alpha
+(q-4)(1+\alpha)
+(q-p)\frac{4\beta}{1+\alpha}\right].
\end{equation}
we can write this constraint in the form
\begin{equation}
\label{minimum}
\frac{q}{2}-2-\alpha+2\beta\frac{q-p}{\alpha+1}>0.
\end{equation}

Similarly, we must require that there is no metastable minimum before
$\sigma_c$, or otherwise the inflaton would become trapped there. This
condition translates to
\begin{equation}
\label{equ:metastable}
\delta_{\rm MS}\equiv\left(\frac{m}{g}\right)^{q-p}
\frac{\kappa_q}{\lambda_p}=\left(\frac{\alpha}{\alpha+1}\right)^{\frac{q-p}{2}}
\left[\frac{\alpha+1}{4\beta}+1\right]
<1.
\end{equation}

Now, initially, the potential (\ref{potential}) has five
free parameters, but if we assume that the Higgs mass is $115$~GeV
\cite{Acciarri:2000ke},
it fixes $\lambda\approx 0.1$, and
also $\kappa_q$ via Eq.~(\ref{equ:kappa}). On the other hand,
Eq.~(\ref{equ:lambda}) fixes $\lambda_p$ from the COBE
data. Therefore, once we choose the integers $p$ and $q$,  only
two free parameters remain, and we may plot our constraints in a
two-dimensional plot.
These constraints are quite complicated and therefore a complete map of
the allowed regions of parameter space must be obtained
numerically. However, it is useful to study the constraints in the limits of small and
large $g$ and $m^2$:
\begin{itemize}
\item{\underline{\bf{$g\rightarrow 0$}} \\
\noindent
In this case, $V_0$ can only depend on $g$ via $\beta$, and
we assume that $\beta$ becomes large at small $g$. Then
\begin{equation}
V_0\sim\beta\sim\lambda_pg^{-p} \ .
\end{equation}
Eq.~(\ref{equ:lambda}) then tells us that
$\lambda_p$ satisfies
\begin{equation}
\lambda_p\sim V_0^{(4-p)/2}\sim
\lambda_p^{(4-p)/2}g^{-p(4-p)/2} \ ,
\end{equation}
which implies that
\begin{equation}
\lambda_p\sim g^{p(p-4)/(p-2)} \ \ \ \ \ \mbox{and} \ \ \ \ \ 
\beta\sim g^{-2p/(p-2)},
\end{equation}
justifying our assumption that $\beta$ diverges at small $g$.
Thus we find
\begin{equation}
V_0\sim g^{-2p/(p-2)} \ ,
\end{equation}
which means that very small
values of $g$ are ruled out because the energy density is too high.
In the same way, we also find
\begin{equation}
\delta_{\rm loop}\sim g^{-4(p-3)/(p-2)} \ ,
\end{equation}
showing that the loop correction also becomes
important in this limit, and therefore our analysis does not apply.
On the other hand, $\delta_{\rm MS} \sim~$constant,
which means that
there is no metastable minimum before $\sigma_c$, and
$V''\sim g^{-4/(p-2)}$, which means that
$\sigma_{\rm min}$ remains a minimum in this limit.
}
\item{\underline{\bf{$g\rightarrow\infty:$}}\\
\noindent
In this limit, we first assume that
$\beta$ becomes small and therefore $V_0$ becomes
independent of $g$. Eq.~(\ref{equ:lambda}) shows that $\lambda_p$ approaches a
constant as well, and therefore $\beta\sim g^{-p}$. This justifies our
assumption, and therefore the energy density behaves well in this
limit. The same is true for the condition (\ref{equ:Vdoubleprime}) because
$V''\sim g^2$. However, $\delta_{\rm
MS}\sim\beta^{-1}\sim g^p$ diverges . Moreover,
$\delta_{\rm loop}\sim g^6$, and therefore the
loop corrections ruin our analysis in this limit.
}
\item{\underline{\bf{$m^2\rightarrow0:$}}\\
\noindent
This is a well-behaved limit in which we
find $\beta\sim~$constant,
$V_0\sim~$constant and $V''\sim~$constant.
Moreover, $\delta_{\rm MS}\sim m^{q-p}$ vanishes, satisfying
Eq.~(\ref{equ:metastable}).
However, because $\delta_{\rm
loop}\sim m^{-2}$, the loop corrections become important,
implying that our analysis breaks down.
}
\item{\underline{\bf{$m^2\rightarrow \infty$}}\\
\noindent
In this limit, we find
\begin{equation}
V_0\sim \beta\sim m^{2p/(p-2)} \ ,
\end{equation}
indicating that the
energy density eventually becomes too high. Similarly,
\begin{equation}
\delta_{\rm loop}\sim m^{4/(p-2)} \ ,
\end{equation}
and the loop corrections become more important. The condition
(\ref{equ:metastable}) is satisfied, because $\delta_{\rm MS}\rightarrow 1$
from below, but
(\ref{equ:Vdoubleprime})
breaks down, because $V''\sim -m^2$ becomes negative at large $m^2$.
}
\end{itemize}
These arguments show that all these extreme limits are either ruled
out or, in the case $m^2\rightarrow 0$,
the loop
corrections get so important that we cannot trust our results. However,
we will show numerically that there is an allowed region of
intermediate parameter values.

\subsection{Baryogenesis from Preheating}

For the mechanism discussed in Section~\ref{ssect:barypre}
to work effectively, it is crucial that the period
of the oscillations is much shorter than the time it takes for the
fields to thermalize. Otherwise, the system would remain near
equilibrium, and it would be impossible to have a non-thermal power
spectrum. On the other hand, the frequency cannot be too high, or
short-wavelength modes of the standard model fields are also excited.
The period of the oscillation is approximately
\begin{equation}
\label{equ:period}
t_{\rm osc}=2\int_{\sigma_a}^{\sigma_b}d\sigma\left(\frac{d\sigma}
{dt}\right)^{-1}=
\sqrt{2}\int_{\sigma_a}^{\sigma_b}\frac{d\sigma}{\sqrt{V_0-V(\sigma)}},
\end{equation}
when $\sigma$ oscillates between $\sigma_a$ and $\sigma_b$. For small
amplitudes, the period is simply given by
\begin{equation}
t_{\rm osc}=\frac{2\pi}{V''(\sigma_{\rm min})} \ ,
\end{equation}
where $V''$ is given by Eq.~(\ref{equ:Vdoubleprime}).

We will first study the special case $V''\approx m_H^2$, $V_0\approx
(250~{\rm GeV})^4$, where we have chosen both the energy density and the
frequency sufficiently high (but not too high) for our mechanism to be
effective. These two conditions fix
the remaining parameters, and in the case $p=5$, $q=6$ we find
$g=0.163$ and $m^2=71000$~GeV$^2$. Let us now study these parameter
values in more detail. Eq.~(\ref{equ:lambda}) fixes $\lambda_5\approx
9.5$~TeV$^{-2}$ and Eq.~(\ref{equ:kappa}) consequently yields
$\kappa_6\approx 5.7$~TeV$^{-3}$.
These imply $\alpha\approx 10.7$ and $\beta\approx 79.0$.
From Eq.~(\ref{sigmacobebound}), we then find $\sigma_{\rm
COBE}\approx 8.8$~eV, which means that, for these parameter values,
the COBE fluctuations were generated well before
$\sigma_c\approx 1.6$~TeV, and therefore the loop corrections are very
small, $\delta_{\rm loop}\approx 10^{-12}$.

\begin{figure}
\center
\epsfig{file=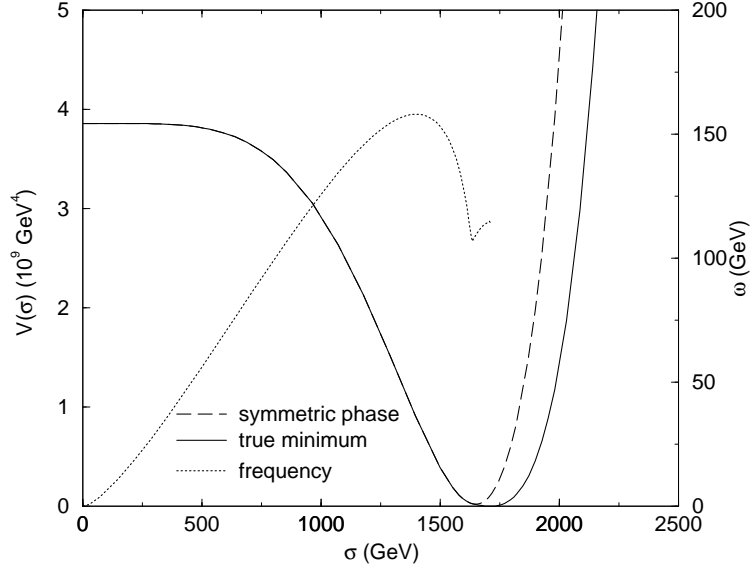,width=10cm}
\caption{
\label{fig:pot}
The inflaton potential at $g=0.163$ and $m^2=71000$~GeV$^2$. The solid
line shows the true minimum at each $\sigma$ and the dashed line shows
the saddle point that corresponds to $\phi=0$.
The dotted line shows the
frequency of the inflaton oscillations as a function of
$\sigma_a$.
}
\end{figure}

The potential is plotted in Fig.~\ref{fig:pot}, where the solid line shows
the effective potential (\ref{equ:effpot}) and the dashed line the symmetric
phase ($\phi=0$) potential.
The minimum is at $\sigma_{\rm min}\approx 1.7$~TeV. As can be
seen from the plot, the potential is extremely flat near $\sigma=0$, and
the Higgs field becomes unstable and the electroweak
symmetry breaks down only very close to the minimum. The amplitude
of the inflaton oscillations is initially much larger than $\sigma_{\rm min}$,
and therefore the SU(2) symmetry gets broken and restored once in every
oscillation. This results in highly efficient excitation of the long-wavelength
modes of the standard model fields .

More quantitatively, we may calculate numerically the period of the
oscillations using Eq.~(\ref{equ:period}). The resulting frequency
$\omega=2\pi/t_{\rm osc}$
as a function
of $\sigma_a$, the minimal value of $\sigma$ reached during the oscillations,
is shown in Fig.~\ref{fig:pot} as a dotted line.
Even when the amplitude is large the
frequency is of order 10 GeV, and therefore the oscillations are
clearly capable of creating a highly non-equilibrium power spectrum for the
standard model fields.

\begin{figure}
\center
\begin{tabular}{ll}
a)&b)\cr
\epsfig{file=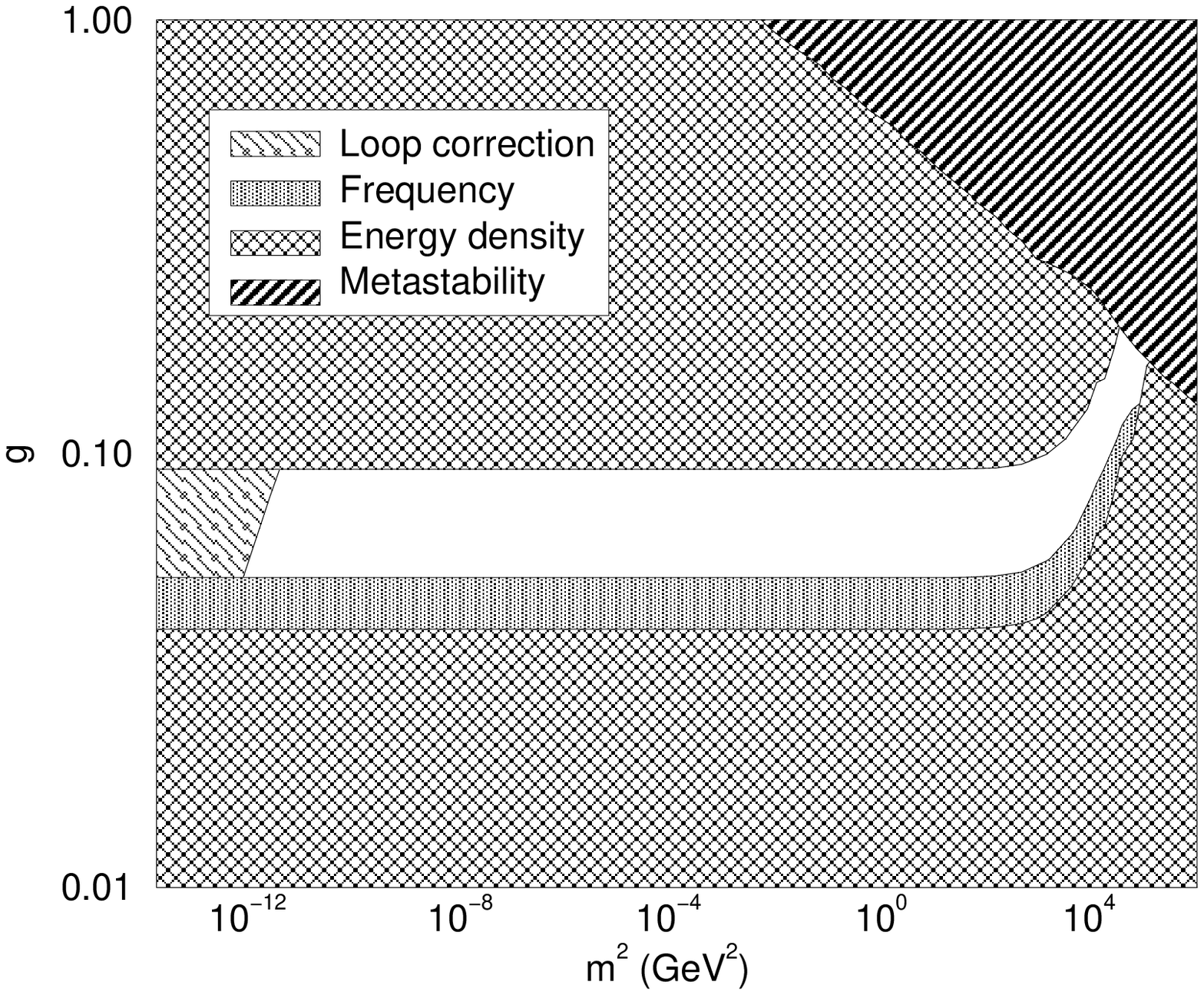,width=8cm}&
\epsfig{file=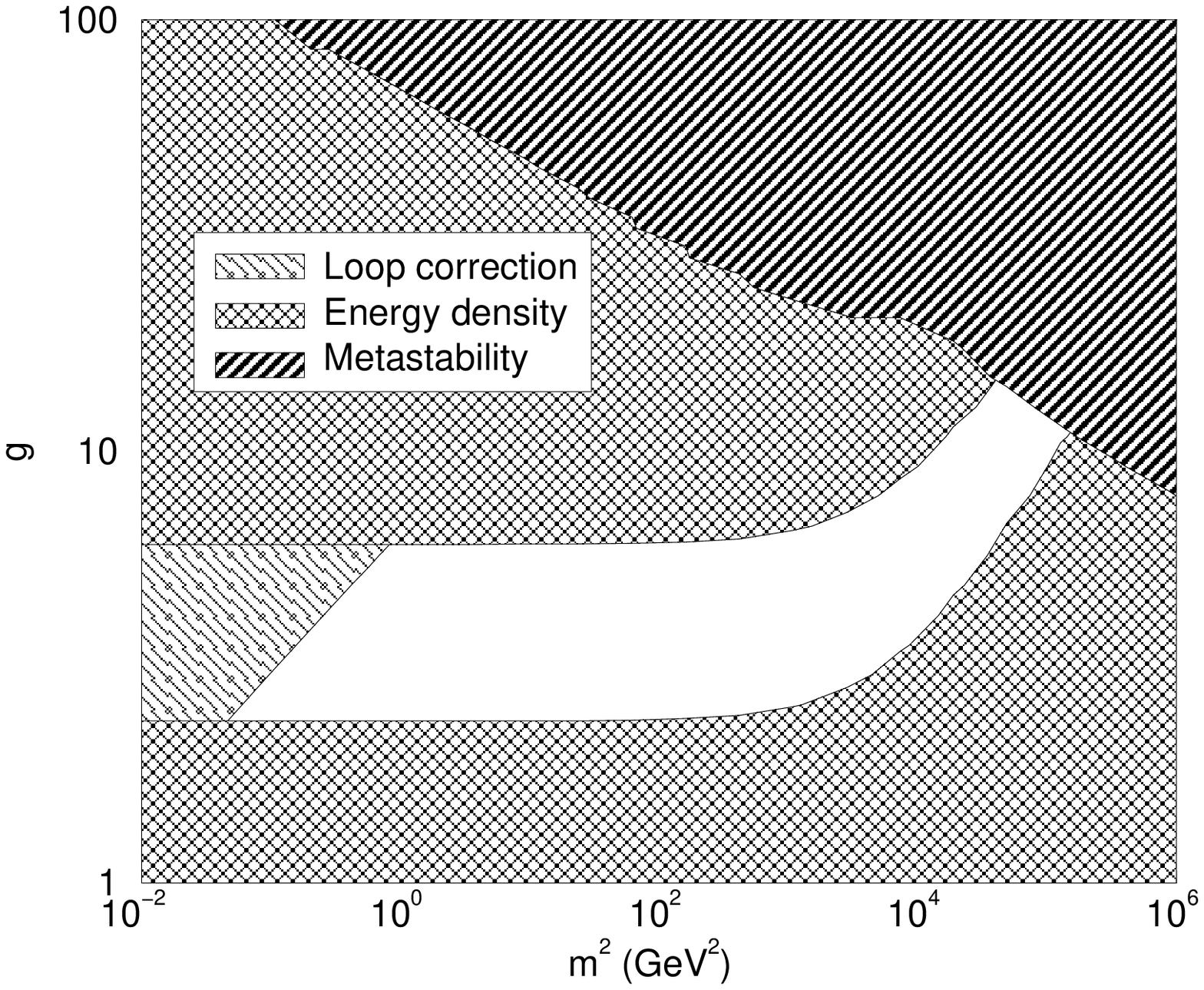,width=8cm}
\end{tabular}
\caption{
\label{fig:exclusion}
The plot of the allowed region of the parameter space.
a) $p=5$, $q=6$, b) $p=6$,
$q=8$. In this case, we haven't plotted the frequency contours,
because the frequency is almost everywhere very high.
}
\end{figure}

In order to see how much fine tuning of parameters is needed, we
show in Fig.~\ref{fig:exclusion}a
the allowed region in $(m^2,g)$ space. The white region in the plot
shows the parameter values than remain allowed after we have excluded
the regions ruled out by
the metastability constraint (\ref{equ:metastable})
the frequency constraint
\begin{equation}
(50~{\rm GeV})^2 < V''(\sigma_{\rm min}) < (200~{\rm GeV})^2 \ ,
\end{equation}
the energy density constraint
\begin{equation}
(150~{\rm GeV})^4 < V_0<(300~{\rm GeV})^4 \ ,
\end{equation}
and the loop correction constraint [see Eq.~(\ref{equ:loopconstraint})]
\begin{equation}
\delta_{\rm loop}<0.1 \ .
\end{equation}
Of particular note is that around the
value $g\approx 0.07$, there is a wide range of allowed values,
\begin{equation}
(1~{\rm keV})^2 \lsim m^2 \lsim (300~{\rm GeV})^2 \ .
\end{equation}

We have repeated the analysis for other values of the parameters $p$ and $q$, with
the result that the constraints become stronger. A natural choice would be $p=6$ and
$q=8$, because the potential would then have a $Z_2$ symmetry. However, in that case the
energy density is typically very high, unless $g$ is large, as shown
in Fig.~\ref{fig:exclusion}b. In such a situation it is questionable
whether our techniques can be applied at all.
Furthermore, since the frequency of the inflaton
oscillations is also typically very high, the inflaton would
decay to high-momentum Higgs modes instead of exciting only the
lowest modes.

Nevertheless, with $p=5$, $q=6$, the model generates naturally the
necessary conditions for the scenario of
Refs.~\cite{Krauss:1999ng,Garcia-Bellido:1999sv}. The
short-wavelength modes are essentially in vacuum, while the long-wavelength modes
have a high effective temperature. When the system thermalizes, this temperature
decreases, the system undergoes a non-thermal phase transition to
the broken phase~\cite{kofm,Tkachev:1996md,Rajantie:2000fd},
and the baryon asymmetry freezes in.
Because of the non-perturbative,
non-equilibrium nature of the mechanism, reliable estimates for the
generated baryon asymmetry can only be obtained by means of numerical
simulations, as discussed in Ref.~\cite{Rajantie:2000nj}.
As an aside, let us point out that because the potential is much more
highly curved in the $\sigma$ direction than in the $\phi$ direction
(see Fig.~\ref{fig:pot}),
the oscillations of $\sigma$ quickly die away, and only $\phi$ keeps
oscillating, as was assumed in Ref.~\cite{Rajantie:2000nj}.

\subsection{Baryogenesis from the Kibble Mechanism}
If the energy density is lower than is needed for the above scenario,
baryogenesis may still take place via the Kibble mechanism as
discussed in Section~\ref{ssect:barykibble}.
In our model $m^2(\sigma)=m^2-g^2\sigma^2$, and the adiabaticity
condition (\ref{adiabaticity}) implies
\begin{equation}
k_{\rm max}^3 \sim g^2 \sigma_c \dot{\sigma}.
\end{equation}
Using $\sigma_c=m/g$, we find
\begin{equation}
n_{\rm configs}
\sim \xi_{\rm max}^{-3} \sim k_{\rm max}^3 \sim g m \dot{\sigma}.
\end{equation}
Depending on CP violation,
this must be higher than
$0.001m_H^3$ for sufficient baryogenesis~\cite{Krauss:1999ng}.

If we insist that inflation does not end before $\sigma_c$, the
slow-roll condition (\ref{flat1}) requires $\dot{\sigma}\ll
V(\sigma_c)^{1/2}\sim m_H^2$, which may be high enough.
We will, however, relax this requirement and assume
that inflation has ended before $\sigma_c$, whereby
we obtain an estimate for the value of
$\dot{\sigma}$ by neglecting the expansion of the universe and using
conservation of energy.
This yields
\begin{eqnarray}
\dot{\sigma}^2&\sim& V(\sigma_{\rm end})-V(\sigma_c)\sim
V_0-V(\sigma_c)
\nonumber\\&=&
\frac{m_H^4}{\lambda}\left[
\frac{1}{p}\beta\left(\frac{\alpha}{\alpha+1}\right)^{p/2}
-\frac{1}{q}\left(\frac{\alpha+1}{4}+\beta\right)
\left(\frac{\alpha}{\alpha+1}\right)^{q/2}
\right].
\end{eqnarray}

In general $\sigma$ will roll past its minimum, reach a
maximum value and turn back. If its initial speed is too high and
there is not enough friction, it may cross $\sigma_c$ again and
restore the electroweak symmetry.
To avoid this, we require that the energy density is not much higher
than is needed for restoring the SU(2) symmetry. In practice, we
choose $V_0<m_H^4/4\lambda$. Note, however, that even higher energy
densities could be acceptable, because lattice
simulations~\cite{Rajantie:2000nj}
suggest that the Higgs winding changes very reluctantly during
symmetry restoration.

The other relevant criteria are the
absence of a metastable minimum (\ref{equ:metastable}) and that the
generated number density of winding
configurations is high enough. The white
region in Fig.~\ref{fig:pot_kibble}a shows the resulting
allowed region of parameter space,
and is centred around
$g\approx 0.2$ and $m^2\approx 1000$~GeV$^2$.

\begin{figure}
\center
\begin{tabular}{ll}
a)&b)\cr
\epsfig{file=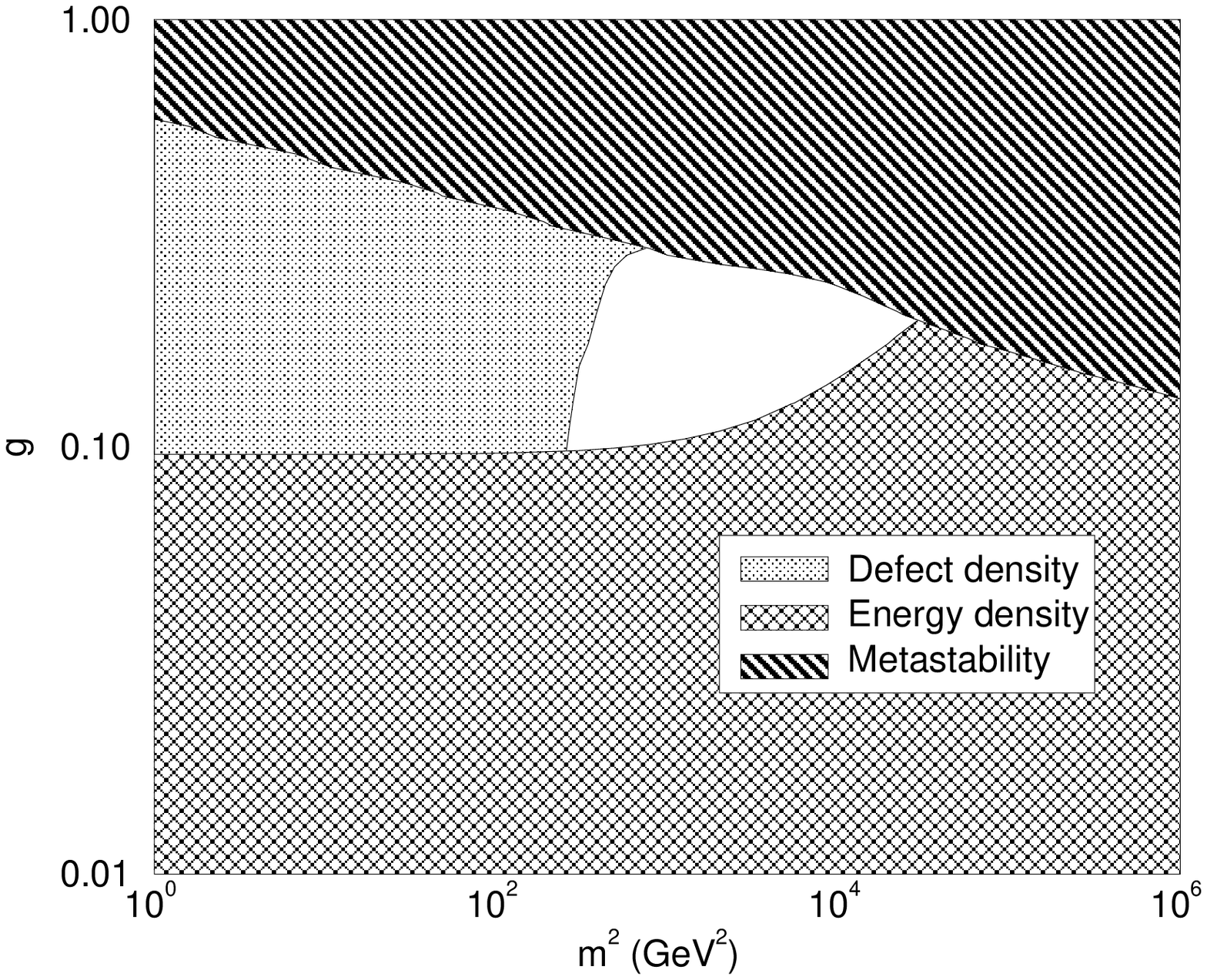,width=8cm}&
\epsfig{file=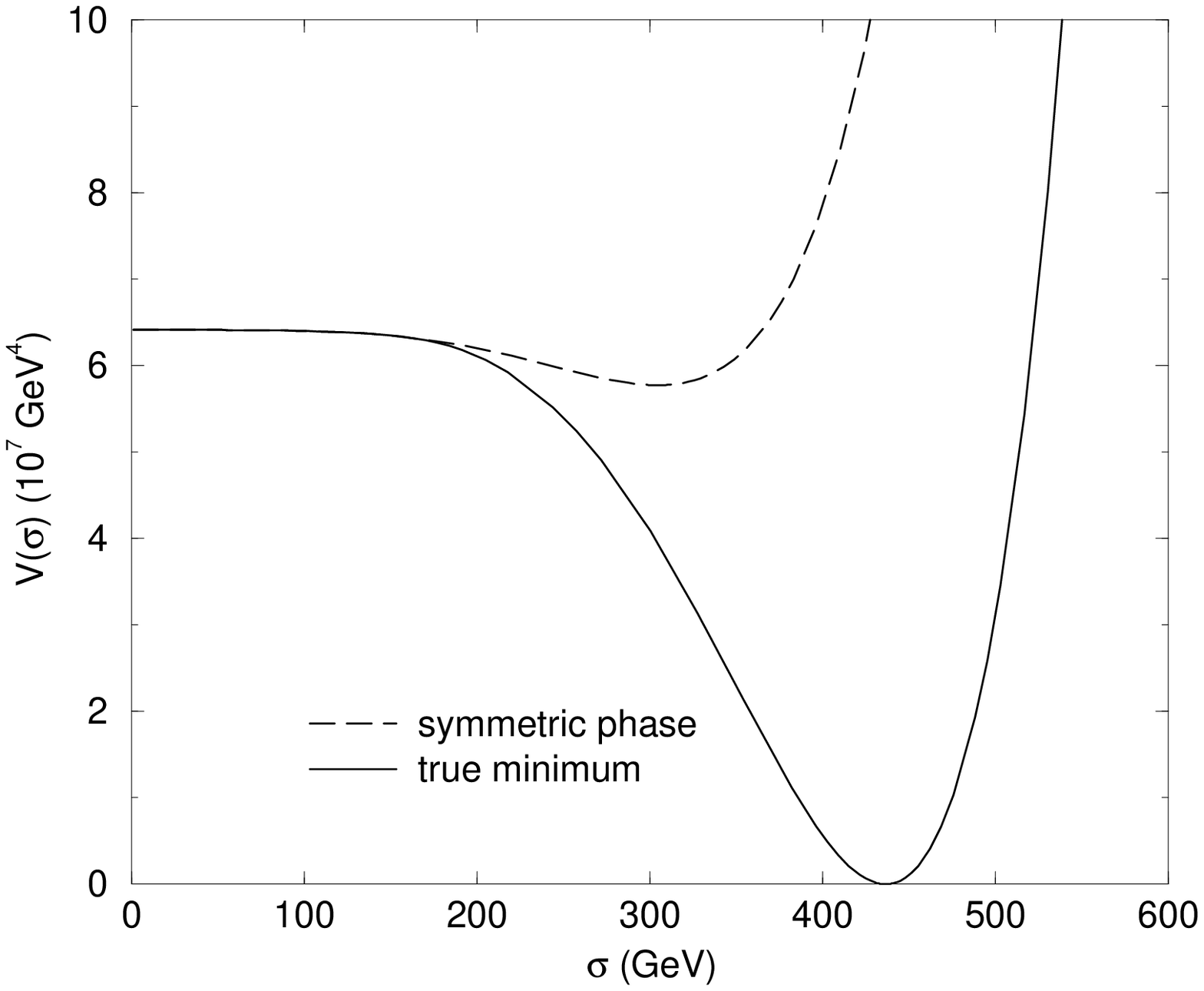,width=8cm}
\end{tabular}
\caption{
\label{fig:pot_kibble}
a) The region of the parameter space where baryogenesis from the Kibble
mechanism is possible.
b) The potential at $g=0.2$, $m^2=1000$~GeV$^2$.
}
\end{figure}

Let us examine these parameters more closely.
The couplings become $\lambda_5\approx 7.3\times
10^{-5}$~GeV$^{-1}$ and $\kappa_6\approx 2.4\times
10^{-7}$~GeV$^{-2}$, and we also find
$\alpha\approx 0.15$, $\beta\approx 0.66$.
The potential is shown in Fig.~\ref{fig:pot_kibble}b. As can be seen from
the plot, even a slight dissipation of energy during the first
oscillation means that the field is no longer able to restore the SU(2)
symmetry, and any Higgs winding number generated is therefore
safe. The speed of the inflaton at $\sigma_c$ is
\begin{equation}
\dot{\sigma}(\sigma_c)\approx 1300~{\rm GeV}^2 \ ,
\end{equation}
and the generated winding number density is
\begin{equation}
n_{\rm configs}\approx 8000~{\rm GeV}^3\approx 0.005m_H^3 \ ,
\end{equation}
which may be sufficient for baryogenesis.
For an accurate estimate of how $n_{\rm configs}$ relates to $n_B/s$,
one would have to solve the SU(2)+Higgs equations of motion in the
presence of a CP violating coupling.

\section{Summary and Conclusions}
\label{conclusions}
In this paper we have addressed the fascinating prospect that it may be
possible to have successful inflation at the TeV scale, in which the
observed fluctuations in the cosmic microwave background are generated.
Following the end of this inflation, the observed baryon asymmetry of the
universe may then be produced. The self-consistent
scenario we have
introduced involves an inverted hybrid inflation model
which allows us to absorb into the renormalization counterterms of the
couplings
the potentially dangerous loop corrections present in ordinary hybrid models
operating at this low energy scale \cite{Lyth:1999ty}. The inflaton is
directly coupled to the standard model Higgs and during inflation the
electroweak symmetry is unbroken. Two possible mechanisms may operate to generate
the baryon asymmetry.
The first is based on the effective non-thermal
restoration of the electroweak symmetry
during preheating, which occurs at the end of the inflationary period.
During symmetry restoration, the Higgs winding and Chern-Simons
numbers can change, and when the system thermalizes, they freeze into
a non-zero value. In this mechanism the baryon washout problem of the
standard electroweak phase transition is avoided, because the
effective cooling rate is determined by the thermalization rate of the
standard model fields rather than the expansion of the universe.

The second possibility is through the zero-temperature Kibble
mechanism occurring at the end of inflation, when the
Higgs field becomes unstable triggering a phase transition.
There is no baryon washout problem because the process takes place at
zero temperature.
The subsequent
evolution of the Higgs field leads to configurations being formed in which
there is a non-trivial Higgs winding number. These unstable configurations
subsequently decay, leading to anomalous fermion number production. Coupled
with the fact that the process is out of equilibrium, and assuming there
exist CP-violating
effects as the configurations unwind, the ingredients are present for
the generation of a baryon asymmetry.

In both cases
we have presented details showing the degree of fine tuning required
to satisfy constraints from both field theory and cosmology.
Not surprisingly there
is certainly some fine tuning required in these models. For example, it turns out that
we require $p=5$ in order to satisfy all the constraints.

It remains to be seen whether such a paradigm is realized in a full,
particle physics inspired model of inflation, but it is certainly of potential importance
that it appears possible to use the same model to generate both the large scale features of
our Universe and the observed baryon asymmetry.

\section*{Acknowledgements}
We are grateful to Paul Saffin for useful
discussions, and Andrei Linde and Gary Felder for useful
correspondence. 
M.T. was supported in part by funds
provided by Syracuse University. A.R. was supported by PPARC and
partially by the University of Helsinki.

\end{document}